\documentclass[aps,prl,twocolumn,superscriptaddress]{revtex4}
\usepackage{graphicx}
\usepackage{bm}
\usepackage{epsfig} 
\usepackage{amsfonts}
\usepackage{amsmath}

\allowdisplaybreaks[1] 

\newcommand*{\be}{\begin{equation}}
\newcommand*{\ee}{\end{equation}}
\newcommand*{\bea}{\begin{eqnarray}}
\newcommand*{\eea}{\end{eqnarray}}

\begin{document}


\title{Tunneling spectra of layered strongly correlated $d$-wave 
superconductors}

\author{Tiago C. Ribeiro}
\affiliation{Department of Physics, University of California, Berkeley, California 94720, USA}
\affiliation{Material Sciences Division, Lawrence Berkeley National Laboratory, Berkeley, California 94720, USA}
\author{Xiao-Gang Wen}
\affiliation{Department of Physics, Massachusetts Institute of Technology, Cambridge, Massachusetts 02139, USA}

\date{\today}

\begin{abstract}
Tunneling conductance experiments on cuprate superconductors
exhibit a large diversity of spectra
that appear in different nano-sized regions of inhomogeneous samples.
In this letter, we use a mean-field approach to the $tt't''J$ model in order
to address the features in these spectra that deviate from the
BCS paradigm, namely, the bias sign asymmetry at high bias, the generic lack 
of evidence for the Van Hove singularity, and the absence of coherence peaks 
at low dopings. 
We conclude that these features can be reproduced in homogeneous
layered $d$-wave superconductors solely due to a proximate Mott insulating
transition.
We also establish the connection between the above tunneling spectral 
features and the strong renormalization of the electron dispersion around 
$(0,\pi)$ and $(\pi,0)$ and the momentum space anisotropy of electronic 
states observed in ARPES experiments.
\end{abstract}


\maketitle

\textit{Introduction} -- 
Tunneling and angle-resolved photoemission spectroscopy (ARPES) are
unique experimental techniques in that they probe the single electron
microscopic physics of strongly correlated systems like the cuprate
$d$-wave superconductors (dSC).
Even though both techniques support the presence of well 
defined Bogoliubov nodal quasiparticles in these materials, 
away from the nodes deviations from the BCS paradigm are encountered
\cite{MS0302,HM0249,PO0182,HF0104,ML0505,HL0401,ZY0401,RS0301,YZ0301}.
Simultaneously describing these deviations in ARPES and tunneling 
experiments is of utmost importance to understand the nature of the
underlying strong local correlations.

In this letter, we focus on the differential conductance
measured by scanning tunneling microscopy (STM) experiments on the
cuprates.
In particular, we address three key aspects observed in the 
data as the electron concentration is increased toward half-filling, namely,
\textit{(i)} an increasing asymmetry of spectra for large positive and 
negative bias;
\textit{(ii)} the ubiquitous absence of peaks caused by the underlying 
Van Hove (VH) singularities in the quasiparticle density of states (DOS);
\textit{(iii)} the gradual depletion of the superconducting (SC) coherence 
peaks.

First, we argue that BCS expectations for dSC are at odds
with the above features.
BCS theory generally predicts that tunneling spectra of quasi-2D
SC materials display VH singularity peaks in addition to two SC 
coherence peaks \cite{QK8801,FK9342}.
However, these VH peaks are not observed by experiments in different 
cuprate families 
(Bi2212 \cite{PO0182,ML0505,MZ9918,OH9735,OM0039}, Bi2201 \cite{KF0111},
LSCO \cite{OM0039}, YBCO \cite{MR9554}). 
Notably, even the SC coherence peaks are absent in certain tunneling 
spectra  \cite{PO0182,HF0104,ML0505,HL0401,KI0404}.

We then show that the aforementioned non-trivial spectral features are 
reproduced in homogeneous SC systems 
close to the Mott insulating state.
Specifically, we use a mean-field (MF) theory of doped Mott insulators
\cite{RW0501,RW_mf} that accounts for many fingerprints of strong 
correlation physics in ARPES data, including the strong band 
renormalization close to $(0,\pi)$ and $(\pi,0)$ \cite{KS9498} and the 
sharp momentum space anisotropy of electronic states, known as the 
nodal-antinodal dichotomy \cite{ZY0401,RS0301,YZ0301}, which we show are 
intimately connected to STM data.
We thus obtain a consistent theoretical microscopic description of ARPES and 
tunneling spectroscopy.
Moreover, we explicitly show that the above deviations from BCS spectra 
follow from how the local electron Coulomb repulsion and the resulting 
short-range antiferromagnetic (AF) correlations affect the phenomenology 
of dSC.

\textit{Tunneling spectra of BCS $d$-wave superconductors} --
We consider the tunneling across a normal-metal -- insulator -- 
superconductor junction in the case of a 
2D superconductor. 
Assuming specular transmission across a thin planar 
junction within the elastic channel, the differential tunneling 
conductance perpendicular to the 2D layers can be 
essentially equated to the quasiparticle DOS
\cite{BD6779,WT9850} and, at zero temperature, it is given by
\cite{M90}
\be
\frac{dI}{dV} = 4e\pi M^2 N(E_F) \sum_{\bm{k}} A(V,\bm{k})
\label{stm:eq:diffcond}
\ee
where $M$ is the tunneling matrix element, $N(E_F)$ is the 
normal-metal DOS, $V$ is the bias of the SC sample with
respect to the metal and $A(\omega,\bm{k})$ is the electron spectral
function in the SC sample. 
The sum in \eqref{stm:eq:diffcond} is over the 2D layer's Brillouin zone.

\begin{figure}
\begin{center}
\includegraphics[width=0.48\textwidth]{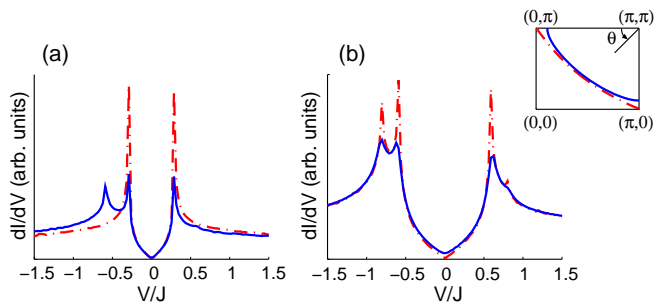}
\caption{\label{stm:fig:bcs}
BCS tunneling differential conductance for $t'=-2t''=-0.12t$ and:
(a) $\Delta = 0.1t$ with $x=0.20$ 
(dash-dot line) and $x=0.10$  
(solid line);
(b) $\Delta = 0.2t, x=0.10$ with (solid line) and
without (dash-dot line) Lorentzian broadening $\Gamma = 0.01t$.
Bias is in units of $J=t/3$.
Plots are obtained for a $2000\times2000$ mesh of the Brillouin zone and
energy resolution of $0.0067 t$.
Inset: Fermi surface of $\varepsilon_{\bm{k}}$ with $t'=-2t''=-0.12t$ for 
$x=0.20$ (dash-dot line) and $x=0.10$ (solid line).}
\end{center}
\end{figure}

In Fig. \ref{stm:fig:bcs} we use \eqref{stm:eq:diffcond} together with
the well known results for the BCS electron spectral function \cite{M90} 
to plot the BCS tunneling differential conductance.
In particular, we take the normal state dispersion $\varepsilon_{\bm{k}}$ 
determined by the $1^{st}$, $2^{nd}$ and $3^{rd}$ nearest-neighbor (NN) 
hopping parameters $t$, $t'$ and $t''$ with $t'=-2t''=-0.12t$ \cite{HOP} so 
that the normal state Fermi surface lies close to $(0,\pi)$ when the electron 
density is $1-x = 0.80$ (inset of Fig. \ref{stm:fig:bcs}), as observed in 
LSCO \cite{IK0204}, Bi2201 \cite{TY0129} and in the antibonding sheet of 
Bi2212 \cite{BL0105}.
In addition, we consider the $d$-wave electron pairing gap function 
$\Delta_{\bm{k}} \equiv \Delta \cos(2\theta)$ with $\Delta = 0.1t$.
The resulting $dI/dV$ spectrum [see dash-dot line in 
Fig. \ref{stm:fig:bcs}(a)] depicts two sharp 
coherence peaks at $V = \pm \Delta$ due to the  SC quasiparticle energy 
dispersion $E_{\bm{k}} = [\varepsilon_{\bm{k}}^2+\Delta_{\bm{k}}^2]^{1/2}$
saddle-point at the SC gap energy.
This behavior is consistent with experimental spectra around optimal 
doping \cite{PO0182,HF0104,ML0505,MZ9918,OH9735,OM0039}.

However, if the electron concentration is increased to $1-x=0.90$ the
normal state Fermi surface intersects the $(0,\pi)-(\pi,\pi)$ line
at a point, which we denote by $\bm{k}_{A}$, that moves away from 
$(0,\pi)$ (inset of Fig. \ref{stm:fig:bcs}).
In this case, the VH singularity in $\varepsilon_{\bm{k}}$ at $(0,\pi)$ and 
$(\pi,0)$ shows up as a separate peak in the BCS tunneling spectrum
at $V= -E_{VH}\equiv - [\varepsilon_{\bm{k}=(0,\pi)}^2+\Delta^2]^{1/2}$
[Fig. \ref{stm:fig:bcs}(a)].
This additional peak appears behind the negative bias SC coherence peak 
unless $\bm{k}_{A} \approx (0,\pi)$ or  $E_{VH} - \Delta$ is so small 
that spectral broadening smears the VH and the SC coherence peaks into a 
single feature.
However, as shown in Fig. \ref{stm:fig:bcs}(b), 
the above double peak structure still survives if, as expected for real
materials upon increasing electron concentration, the $d$-wave gap is 
doubled to $\Delta = 0.2t$ and a broadening width $\Gamma = 0.01t$ 
is included \cite{BROADENING}.

\begin{figure}
\begin{center}
\includegraphics[width=0.48\textwidth]{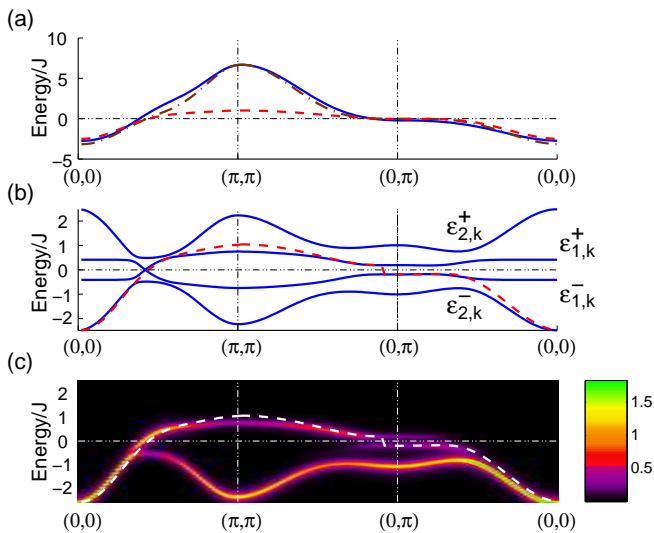}
\caption{\label{stm:fig:bands}
Energy dispersions along $(0,0)-(\pi,\pi)-(0,\pi)-(0,0)$.
(a) Normal state dispersions using parameters from Ref. \cite{NR9515}
(solid line), Ref. \onlinecite{N0051} (dash-dot line) and 
$\epsilon_{\bm{k}}^N$ defined in main text (dashed line).
(b) MF dispersions $\epsilon_{1,\bm{k}}^{\pm}$ and 
$\epsilon_{2,\bm{k}}^{\pm}$ for $x=0.18$ and parameters used in Ref.
\onlinecite{RW0501}.
(c) MF spectral function with Lorentzian broadening $\Gamma = J/5$ 
where $J=150\text{meV}$.
Dashed line in (b) and (c) plots
$[(\epsilon_{\bm{k}}^N)^2+\Delta_{\bm{k}}^2]^{1/2} \times 
\text{sgn}(\epsilon_{\bm{k}}^N)$.
}
\end{center}
\end{figure}

This behavior contrasts with the experimental observation that no such
double peak structure occurs as the hole doping level is 
decreased \cite{PO0182,ML0505,MZ9918,OH9735,OM0039} 
and implies that the actual
dispersion along $(0,\pi)-(\pi,\pi)$ in the cuprate materials is of the 
order of the SC gap energy
scale, namely $\sim 0.1t$, which is an order of magnitude lower than $t$.
Such a flat dispersion in the antinodal region is supported
by ARPES data in Bi2201, Bi2212 and YBCO \cite{KS9498}.
Since ARPES observations also constrain the much larger nodal dispersion 
energy scale, as well as the curvature of the Fermi surface, several 
fine-tuned phenomenological parameters are required to fit the observed 
normal state dispersions 
[Fig. \ref{stm:fig:bands}(a) displays the dispersions obtained by
Norman \textit{et al.} from experimental fits involving up to five hopping 
terms \cite{NR9515,N0051}].
As Ref. \onlinecite{DN9428} points out, the ubiquitous
discrepancy between the energy scale of the nodal dispersion and 
that of the extended flat region around $(0,\pi)$ throughout the
various cuprate families and doping levels suggests
that interactions strongly renormalize the electron dispersion.
Below, we argue that such phenomenological dispersions are 
natural to doped Mott insulators.

\textit{Renormalization of single electron properties} --
To discuss how the single electron dispersion relation and the momentum 
space distribution of spectral weight are renormalized in doped Mott 
insulators we consider the $tt't''J$ model, whose numerical calculations 
reproduce many spectral features of the cuprates \cite{TM0017,T0417}.

In the half-filled AF insulator, the
electron dispersion is renormalized by the strong local AF correlations.
In particular, NN hopping is heavily frustrated by the AF background
and, for $t'=t''=0$, holes hop coherently within the same sublattice by
virtue of spin fluctuations \cite{KL8980}. 
As a result, the width of the nodal dispersion is reduced by a 
factor of $\sim 10$ from the bare value $8t$ down to $2.2J$
\cite{TM0017}.
$t' \approx -2t''$ control the dispersion along $(0,\pi)-(\pi,0)$ and,
since they describe intrasublattice hopping processes which 
are not frustrated by AF correlations, the dispersion width 
along this line is only renormalized by a factor $\sim 2-3$ \cite{TM0017}.

Away from half-filling, holes distort the AF spin background and change 
the surrounding spin configuration in order to
coherently hop between NN sites and thus gain extra kinetic energy.
Interestingly, exact diagonalization studies show that the resulting
local spin correlations strongly renormalize $t'$ and $t''$ \cite{R0402}.
This effect, which is predicted in spin liquids \cite{RW0301},
is consistent with the experimentally observed doping induced 
flatness of the dispersion in the antinodal region \cite{KS9498}
since the dispersion along $(0,\pi)-(\pi,0)$ is controlled by 
$t'$ and $t''$.

The ``doped carrier'' MF approach introduced in 
Refs. \onlinecite{RW0501,RW_mf} accounts for the above interplay between the 
exchange energy of localized spins and the kinetic energy of delocalized holes.
Here, we present only a qualitative description of the MF theory and 
defer the readers to the above references for formal details.
This theory describes doped Mott insulators in terms of two different 
fermions:
(i) ``dopons'', which have the same charge $+e$ and spin-$1/2$ as holes, 
describe vacancies surrounded by an AF spin configuration;
(ii) ``spinons'', which have no electric charge, describe spin-$1/2$
excitations of the spin background.
The MF Hamiltonian thus has four fermionic bands:
$\epsilon_{1,\bm{k}}^-=-\epsilon_{1,\bm{k}}^+$ and 
$\epsilon_{2,\bm{k}}^-=-\epsilon_{2,\bm{k}}^+$ [Fig. \ref{stm:fig:bands}(b)].
If dopons and spinons do not hybridize the high energy bands
$\epsilon_{2,\bm{k}}^{\pm}$ describe the dynamics of holes in an AF background
and, at MF level, the spinon $\epsilon_{1,\bm{k}}^{\pm}$ bands do not 
contribute to electron spectral properties.
The process of hybridizing dopons and spinons,
which describes the change in spin correlations due to
hole hopping, couples
spin and charge dynamics and transfers electron spectral weight from 
$\epsilon_{2,\bm{k}}^{\pm}$ to the lower energy 
$\epsilon_{1,\bm{k}}^{\pm}$ bands [Fig. \ref{stm:fig:bands}(c)].
This spectral weight transfer is not uniform in momentum space and
reflects the dispersion of the dopon band, which has lower energy
at $(\pi/2,\pi/2)$ than at $(0,\pi)$.
As a result, the $\epsilon_{1,\bm{k}}$ band has more spectral weight near 
the nodal points  than in the antinodal  region \cite{RW0501},
in agreement with the nodal-antinodal dichotomy in ARPES data 
\cite{ZY0401,RS0301,YZ0301}.
In conformity with other approaches to the related Hubbard 
model \cite{ST0401}, we conclude that this dichotomy results from 
short-range correlations, namely AF correlations \cite{RW_mf,R0402}, 
in dSC.

The low energy dispersions $\epsilon_{1,\bm{k}}^{\pm}$ derive from the 
spinon bands and, thus, from the spin exchange interaction.
Therefore, they are controlled by a single energy scale, namely $J$, which
leads to an almost flat dispersion near the antinodal points.
We thus find that the spin background strongly renormalizes 
the electron dispersion in the antinodal region.
To explicitly compare the MF dispersion with the experimental
fits to ARPES data we introduce the underlying normal dispersion
$\epsilon_{\bm{k}}^N$ such that
$[(\epsilon_{\bm{k}}^N)^2+\Delta_{\bm{k}}^2]^{1/2}$ equals:
(i) $\epsilon_{1,\bm{k}}^+$ at $\bm{k}_{A}$, $(0,\pi)$ and at the nodal point;
(ii) $\epsilon_{2,\bm{k}}^+$ at $(0,0)$ 
[Figs. \ref{stm:fig:bands}(b) and \ref{stm:fig:bands}(c)].
In Fig. \ref{stm:fig:bands}(a) we plot $\epsilon_{\bm{k}}^N$ for $x=0.18$
against the normal dispersion fits from Refs. \onlinecite{NR9515,N0051}
along major symmetry directions.
Clearly, below the Fermi level $\epsilon_{\bm{k}}^N$ captures the 
energy scale of the nodal 
dispersion and the flatness around antinodal points.
ARPES does not probe the dispersion above the Fermi energy and both
experimental fits use band calculation results to fix the energy at 
$(\pi,\pi)$ \cite{NR9515}, whence the mismatch between
$\epsilon_{\bm{k}}^N$, obtained within the $tt't''J$ model context, and 
these fits above the Fermi level.
Such a mismatch is supported by exact diagonalization calculations of  
the $tt't''J$ model which indicate that the dispersion above the Fermi 
level is less dispersive than expected from bare hopping parameters
\cite{T0417}.

\begin{figure}
\begin{center}
\includegraphics[width=0.48\textwidth]{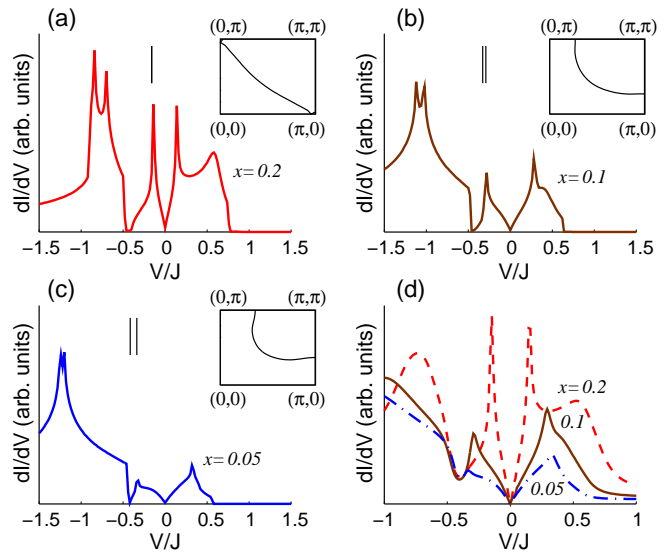}
\caption{\label{stm:fig:versus}
MF tunneling differential conductance for (a) $x=0.20$, (b) $x=0.10$ and 
(c) $x=0.05$. 
(d) Same spectra with broadening given by
$\Sigma''(\omega) = -\omega^2/(5J)$.
Momentum space and energy resolution are the same as in 
Fig. \ref{stm:fig:bcs}.
Insets of (a)-(c) plot minimal gap locus of $\epsilon_{1,\bm{k}}^{\pm}$.
Vertical bars in (a)-(c) denote VH singularity and SC gap energies.
}
\end{center}
\end{figure}

\textit{Tunneling spectra of strongly correlated $d$-wave superconductors} --
We now consider the tunneling conductance for the above MF theory of 
doped Mott insulators.
We use the same parameters as in Ref. \onlinecite{RW0501} 
to obtain the $dI/dV$ curves at $x=0.20$, $x=0.10$ and $x=0.05$ 
[Figs. \ref {stm:fig:versus}(a)-\ref{stm:fig:versus}(c)].
We point out that the MF theory produces a sharp
electron spectral peak at the $\epsilon^\pm_{2,\bm{k}}$ band.
However, we expect this spectral peak to be significantly broadened 
if one goes beyond the MF approximation 
since $\epsilon^\pm_{2,\bm{k}}$ are high energy bands which 
have many channels to decay.
Hence, in order to better compare to experiments, in
Fig. \ref {stm:fig:versus}(d) we depict the above 
spectra with a frequency dependent broadening 
given by $\Sigma''(\omega) = -\omega^2/(5J)$.

Looking at Figs. \ref {stm:fig:versus}(a)-\ref{stm:fig:versus}(c) 
we recognize the separate contribution to the tunneling DOS from bands
$\epsilon_{2,\bm{k}}^-$, $\epsilon_{1,\bm{k}}^-$ and $\epsilon_{1,\bm{k}}^+$.
The positive energy $\epsilon_{2,\bm{k}}^+$ band does not appear since it has
vanishing spectral weight [see Fig. \ref{stm:fig:bands}(c)].
The presence of the negative energy bands $\epsilon_{2,\bm{k}}^-$ and 
$\epsilon_{1,\bm{k}}^-$ leads to a peak-dip-hump in the $dI/dV$ curve as 
observed by experiments \cite{ML0505,MZ9918}.
This negative bias hump, which moves to higher energy as doping is lowered
\cite{ML0505,MZ9918}, derives from the lower Hubbard band and renders 
the tunneling conductance plots highly asymmetric for large 
positive and negative bias.
At smaller values of $x$ the spectral weight in the $\epsilon_{1,\bm{k}}^-$ 
and $\epsilon_{1,\bm{k}}^+$ bands is transfered to the high-energy hump
and, thus, the above bias sign asymmetry is enhanced upon underdoping
as seen in STM data \cite{PO0182,ML0505,HL0401,KI0404}.
In particular, the ratio of integrated spectral weight at positive bias over 
that at negative bias is $2x/(1-x)$, in perfect
agreement with sum rules applicable to the generalized-$tJ$ model 
\cite{RS0501}.
The above behavior is a hallmark of models where strong local Coulomb 
repulsion between electrons makes it easier to add a hole
to the sample than to add an electron. 

Figs. \ref{stm:fig:versus}(a)-\ref{stm:fig:versus}(c) depict the 
doping evolution of the minimum gap locus, which reproduces 
experiments \cite{IK0204}, and denote 
the VH singularity and SC gap energies, $E_{VH}$ and $\Delta$ respectively, 
by the small vertical bars.
Changing from $x=0.20$ to $x=0.10$ the wave vector $\bm{k}_{A}$ distances
from $(0,\pi)$, yet $E_{VH} - \Delta$ remains extremely small
since the dispersion is very flat in the antinodal region.
Therefore, no extra peak develops next to the SC coherence peak, unlike 
the cases depicted in Fig. \ref{stm:fig:bcs}.
The energy difference $E_{VH} - \Delta$ increases for $x=0.05$, yet,
we observe a strong suppression of the SC
coherence peak in the tunneling spectrum of Fig. \ref {stm:fig:versus}(c)
instead of a double peak structure.
Below, we explain this behavior, which is in stark contrast with weak 
coupling BCS predictions.

The curves in Fig. \ref {stm:fig:versus} show how the SC coherence peak
evolves with doping level.
The $x=0.20$ spectrum displays well defined SC coherence peaks, in 
agreement with spectra of samples around optimal doping 
\cite{PO0182,HF0104,ML0505,MZ9918,OH9735,OM0039}.
In the underdoped $x=0.10$ spectrum the SC coherence peaks
are pushed to higher energy due to the SC gap doping dependence
and lose spectral weight, as seen in experiments
\cite{PO0182,ML0505,MZ9918,OH9735}.
The spectrum for the SC state at $x=0.05$ hardly resolves the SC coherence 
peaks as found in deeply underdoped Bi2212 \cite{ML0505,PO0182} and 
NaCCOC \cite{HL0401,KI0404}.
The gradual depletion of the SC coherence peak intensity as the 
half-filled state is approached follows from the correspondingly 
gradual depletion of spectral weight in the antinodal region
which underlies
the formation of the low energy spectral weight arcs observed both 
in ARPES experiments \cite{ZY0401,RS0301,YZ0301} and in theoretical 
approaches \cite{RW0501,ST0401}.
In our calculation, this behavior explicitly results from the
presence of short-range AF correlations in dSC.

As we emphasize in Fig. \ref {stm:fig:versus}(d), the doping evolution 
of the $dI/dV$ curves leads to qualitatively different line shapes around 
optimal doping and in the deeply underdoped regime. 
We refer to Fig. 2 of Ref. \onlinecite{ML0505} to show that the above 
results reproduce experimental evidence for different types of spectra 
in SC cuprate materials.
Since this STM spectral diversity
occurs in samples which are inhomogeneous on the nano-scale 
\cite{PO0182,HF0104,ML0505,KI0404,FC0607,ML0548},
it has been proposed that these non-trivial line shapes are 
intrinsically related to the presence of the underlying sample 
inhomogeneity \cite{FC0607}.
The possible relevance of coexisting orders \cite{ML0505,HL0401} and  phase 
separation scenarios \cite{HF0104,KI0404} has also been pointed out in the 
literature.
However, here we show that the aforementioned unusual and diverse spectra 
can be reproduced in homogeneous SC systems solely due to the presence of 
strong correlations that follow from a proximate Mott insulating state.
This theoretical assertion supports that electron dynamics in these 
materials is, to a large extent, determined by 
properties which are local even when compared to the nano-sized
sample inhomogeneities, in conformity with experimental 
results \cite{ML0548}.

\begin{acknowledgments}
This work was supported by the Funda\c c\~ao 
Calouste Gulbenkian Grant No. 58119 (Portugal), 
by the NSF Grant No. DMR--01--23156,
NSF-MRSEC Grant No. DMR--02--13282 and NFSC Grant No. 10228408
and by the DOE Grant No. DE-AC02-05CH11231.

\end{acknowledgments}

\newcommand*{\PR}[1]{Phys.\ Rev.\ {\textbf {#1}}}
\newcommand*{\PRL}[1]{Phys.\ Rev.\ Lett.\ {\textbf {#1}}}
\newcommand*{\PRB}[1]{Phys.\ Rev.\ B {\textbf {#1}}}
\newcommand*{\PRD}[1]{Phys.\ Rev.\ D {\textbf {#1}}}
\newcommand*{\PTP}[1]{Prog.\ Theor.\ Phys.\ {\textbf {#1}}}
\newcommand*{\MPL}[1]{Mod.\ Phys.\ Lett.\ {\textbf {#1}}}
\newcommand*{\JPC}[1]{Jour.\ Phys.\ C {\textbf {#1}}}
\newcommand*{\RMP}[1]{Rev.\ Mod.\ Phys.\ {\textbf {#1}}}
\newcommand*{\RPP}[1]{Rep.\ Prog.\ Phys.\ {\textbf {#1}}}
\newcommand*{\PHY}[1]{Physics {\textbf {#1}}}
\newcommand*{\ZP}[1]{Z.\ Phys.\ {\textbf {#1}}} 
\newcommand*{\JETP}[1]{Sov.\ Phys.\ JETP Lett.\ {\textbf {#1}}}
\newcommand*{\PLA}[1]{Phys.\ Lett.\ A {\textbf {#1}}}
\newcommand*{\AP}[1]{Adv.\ Phys.\ {\textbf {#1}}}
\newcommand*{\JLTP}[1]{J.\ Low Temp.\ Phys.\ {\textbf {#1}}}
\newcommand*{\SC}[1]{Science\ {\textbf {#1}}}
\newcommand*{\NA}[1]{Nature\ {\textbf {#1}}}
\newcommand*{\CMAT}[1]{cond-mat/{#1}}
\newcommand*{\JPSJ}[1]{J.\ Phys.\ Soc.\ Jpn.\ {\textbf {#1}}}
\newcommand*{\PC}[1]{Physica.\ C {\textbf {#1}}}
\newcommand*{\JPCS}[1]{J.\ Phys.\ Chem.\ Solids\ {\textbf {#1}}}
\newcommand*{\APNY}[1]{Ann.\ Phys.\ (N.Y.) {\textbf {#1}}}
\newcommand*{\ANNA}[1]{Annalen der Physik {\textbf {#1}}}
\newcommand*{\SSC}[1]{Solid State Commun.\ {\textbf {#1}}}
\newcommand*{\SST}[1]{Supercond.\ Sci.\ Technol.\ {\textbf {#1}}}
\newcommand*{\PRPT}[1]{Phys.\ Rep.\ {\textbf {#1}}}
\newcommand*{\JESRP}[1]{J.\ Electron Spectrosc.\ Relat.\ Phenom.\ {\textbf {#1}}}


\begin{thebibliography}{99}
\bibitem{BROADENING} $\Gamma = 0.01t$ is consistent with experiments
\cite{OH9735,HF0104}.
\end{thebibliography}
\end{document}